# Light Streak Photometry and Streaktools

Joshua L. Goodeve
Trottier Space Institute, McGill University, 845 Sherbrooke St W, Montreal, Quebec H3A 0G4, Canada

## Abstract

The accuracy of photometric calibration has gradually become a limiting factor in various fields of astronomy, limiting the scientific output of a host of research. Calibration using artificial light sources in low Earth orbit remains largely unexplored. Here, we demonstrate that photometric calibration using light sources in low Earth orbit is a viable and competitive alternative/complement to current calibration techniques, and explore the associated ideas and basic theory. We present the publicly available Python code *Streaktools* as a means to simulate and perform photometric calibration using real and simulated light streaks. Using *Streaktools*, we perform "pill" aperture photometry on 131 simulated streaks, and Markov chain Monte Carlo based point-spread-function (PSF) model-fitting photometry on 425 simulated streaks in an attempt to recover the magnitude zeropoint of a real exposure of the Dark Energy Camera instrument on the Blanco 4 m telescope. Our results show that calibration using pill photometry is too inaccurate to be useful, but that PSF photometry is able to produce unbiased and accurate ($1\sigma$ error = 3.4 mmag) estimates of the zeropoint of a real image in a realistic scenario, with a reasonable light source. This demonstrates that light-streak photometry is a promising alternative and complement to established techniques, which should be explored and tested further.

*Unified Astronomy Thesaurus concepts:* Calibration (2179); Flux calibration (544); Artificial satellites (68); Astronomy software (1855)

## 1. Introduction

With the great improvement in astronomical instrumentation over the past few decades, accurate photometric calibration is becoming a limiting factor in certain fields of astronomy. The general procedure for calibration is straightforward: a source of known intensity is used to calibrate the brightness scale for an instrument under certain conditions. Certain criteria apply to these calibrators. Astronomical telescopes are focused to infinity, and so a calibration source must be far away; nearby sources would necessitate impractical refocusing of the telescope. This also implies that the sources must be very bright, so as to appear bright from larger distances. Stars and quasars have long served as photometric standards in this manner.

### 1.1. Artificial Sources

It is also possible to use artificial sources for photometric calibration, though the previous requirements still apply (J. Albert 2012). Such an artificial source must be in the upper atmosphere, or in space. Any closer to the ground and the calibration would not be useful; one loses the ability to calibrate for a certain air mass, and observations requiring such calibration are made far above the horizon. The use of artificial light sources for calibration has several advantages. First, light sources can be engineered to very precise specifications and can have their output measured with very high precision on the ground. Second, the output of an artificial source can be monitored in real time using stable, calibrated onboard detectors. There are also disadvantages to this approach. Useful sources placed many tens to hundreds of kilometers away must be reasonably bright, a considerable engineering requirement for satellites or balloon payloads. Another set of disadvantages specific to satellites comes from the large motion they will have against the sky in most orbits. This has consequences that we will explore in more detail in sections to come.

### 1.2. Description of Following Sections

In Section 2 we explore the theory of light-streak-based calibration, where satellites with precise light sources are used for photometric calibration. We describe the situation mathematically with several assumptions and detail how the key object of interest, the *magnitude zero point* of the instrument under a certain set of conditions, may be recovered. We attempt to quantify uncertainties in these measurements by examining the propagation of errors due to engineering tolerances and uncertainty in satellite position. Next, we discuss methods for applying streak photometry analogous to stellar aperture and point-spread-function (PSF) photometry. In the next section, with the theory and methodology laid out, we introduce *Streaktools*, a set of tools implemented in Python for simulating streaks in real images using arbitrary PSFs, and for making measurements using the discussed methods on both simulated and real streaks. Sections 4, 5, and 6 describe our use of *Streaktools* to simulate to several hundred streaks and use them to recover the magnitude zero point of a real image, using several different implemented techniques.

## 2. Theory

### 2.1. Magnitude and Zero Points

A general expression for the apparent magnitude *m* of an object with an integrated flux *F* through some bandpass is

given by

$$m = -2.5 \log_{10}\left(\frac{F}{F_{\text{ref}}}\right) + M_{\text{ref}}, \quad (1)$$

where $F_{\text{ref}}$ is the integrated flux of an object with apparent magnitude $M_{\text{ref}}$. Selection of $F_{\text{ref}}$ and $M_{\text{ref}}$ fixes the relationship between magnitude and flux. In CCD images (with ideally linear detector response), the number of counts $C$ received from a source is proportional to integrated flux, and so one can replace $F$ and $F_{\text{ref}}$ with a measured number of counts $C$ and some reference number of counts $C_{\text{ref}}$, which has a known magnitude $M_{\text{ref}}$. Convention is then to set $C_{\text{ref}}$ to one, and call $M_{\text{ref}}$ the magnitude zero-point $M_0$; the magnitude corresponding to a single count in the image. The relation between magnitude and CCD image counts ($C$) in ADU is then

$$m = -2.5 \log_{10}(C) + M_0. \quad (2)$$

The goal of photometric calibration is to accurately measure $M_0$ so that (2) can be used to calculate all other magnitudes. $M_0$ can be calculated from (2) if one can know both the number of counts $C$ and associated magnitude $m$ of any source. Photometric calibration then consists of calculating or otherwise establishing the magnitude $m$ of a calibrator, and measuring the number of counts $C$ associated with the source in an image. We presently discuss the calculation of the magnitude associated with a light streak in an image.

### 2.2. Calculating Streak Magnitudes

Note that here we will be working in the AB magnitude system, the definition of which lends itself to the following calculation. $m$ can be calculated from (1), since we can calculate the expected flux $F$ of the source, and the definition of the AB system specifies $F_{\text{ref}}$ and $M_{\text{ref}}$ for us. Note that here, by integrated flux we mean the total power per unit area of telescope aperture that makes it through the bandpass (effectively this is the flux from frequencies that the instrument is sensitive to, by design), whereas irradiance $I$ refers to the power per unit area provided by the source at the telescope aperture, which the reader may know as *radiant flux*. This nomenclature has been chosen to avoid confusion between the two. We further assume that this source is effectively monochromatic, with a specific and well-known wavelength, as in a laser. The source flux may then be calculated as

$$F = I \cdot S(\lambda_0), \quad (3)$$

where $S(\lambda_0)$ is the bandpass power transmission at the source wavelength $\lambda_0$. This is not, however, the flux that will be measured by the telescope, which measures flux by integrating it over a length of time. Assume now that the streak is caused by a pulse of known *pulse interval* $\Delta t$, and that the exposure time is $T$—the *effective* flux *measured by the instrument* $F_{\text{meas}}$ will now be

$$F_{\text{meas}} = I \cdot S(\lambda_0) \frac{\Delta t}{T}. \quad (4)$$

For example, if the calibration source was only emitting for 1 s during a 3 s exposure, its calculated integrated flux would be 1/3 of its true value, since this is calculated by normalizing by exposure time. In the AB magnitude system, a source is defined to have a magnitude of zero when its spectral-flux density $f_\nu$ is 3631 Jy. A Jansky (Jy) is a unit of spectral-flux density defined as $10^{-26}\,\text{W}\,\text{m}^{-2}\,\text{Hz}^{-1}$. The total integrated flux $F_0$ of a (by definition) zero-magnitude source (AB system) can then be calculated as

$$F_0 = \int S(\lambda') f_\lambda(\lambda') d\lambda' = \int S(\nu')(3631 Jy) d\nu', \quad (5)$$

where $f_\lambda$ is the flux per unit wavelength corresponding to the zero-magnitude spectral-flux density, 3631 Jy, and $S$ is the bandpass transmission profile. Since this reference flux by definition corresponds to a magnitude of zero, (1) then lets us calculate the theoretical magnitude of this source as

$$m = -2.5 \log_{10}\left(\frac{F_{\text{meas}}}{F_0}\right), \quad (6)$$

where $F_0$ is given by Equation (5), and $F_{\text{meas}}$ is given by Equation (4). Note that the accuracy of our approach depends on the knowledge of the filter transmission profile through $F_{\text{meas}}$ and $F_0$. For our purposes we assume that this profile is known with high accuracy, and we also point out that this issue is not unique to the approach to calibration detailed here. Finally, we need an expression for the irradiance of the source at the site of observation, to plug into Equation (3). Here we assume the source to be a laser-based integrating sphere (IS). An integrating sphere works by repeated diffusing reflections of an interior light source off of its inner surface. Letting this diffused light out of a small output port produces a light intensity profile with uniform *radiance R* (radiated power, per unit area perpendicular to the radiation direction, per unit solid angle radiated into), making it an excellent light source for our purposes. We will call the angle between the light source axis and the source-observatory line $\theta$. The irradiance at the telescope can then be calculated using the known integrating sphere radiance $R_{IS}$ and aperture area $A_{IS}$, and the distance to the satellite source, $d$:

$$I(d, \theta) = \frac{R_{IS} A_{IS} \cos(\theta)}{d^2}. \quad (7)$$

The factor of $\cos(\theta)$ appears in Equation (7) because the area of the integrating sphere aperture, projected onto the plane of observation, is smaller than that area viewed on axis. Any appropriate irradiance may be plugged into (3) and the analysis here performed to determine the magnitude corresponding to the streak.

### 2.3. Streak Photometry

In this section, we describe methods for recovering the number of counts $C$ associated with a streak in a CCD image.

#### 2.3.1. Pill Aperture Photometry

The natural extension of the ubiquitous methods of circular and elliptical aperture photometry is *pill aperture photometry*, pioneered by W. Fraser et al. (2016). With this technique, the usual circular apertures are drawn out into pill shapes to encompass long streaks in astronomical images.

An example pill aperture, drawn appropriately around a simulated light-streak, is shown in Figure 1. The pill is characterized by a rectangular center section, with hemispherical end caps. The pill ideally contains only unbiased noise and counts from the streak. The number of counts $C$ associated with the streak is then obtained simply by counting the pixels inside of the aperture. This is computationally straightforward, but has serious drawbacks. Long pill-shaped apertures necessarily cover more sky than for a stationary source with the same flux, reducing the final signal-to-noise ratio. The risk of other sources in the field intersecting the streak and the pill also increases with the length of the streak; these sources add unwanted flux if not source subtracted, and remove flux or at the very least add noise even with good source subtraction. Another issue is that the larger integration area of the pill requires very accurate sky-background subtraction. For example, if after sky subtraction, a residual sky background of just 0.05 counts pixel$^{-1}$ remains, then a pill aperture 1400 pixels long and 30 pixels wide (which can be reasonable, depending on the length of the streaks in question) could contain thousands of extra unexpected counts, which for reasonable light sources is far too much for precise calibration. These issues are severe enough that we end our analysis of this approach here. In Section 5, we detail the use of simulations to show that this approach is unfeasible.

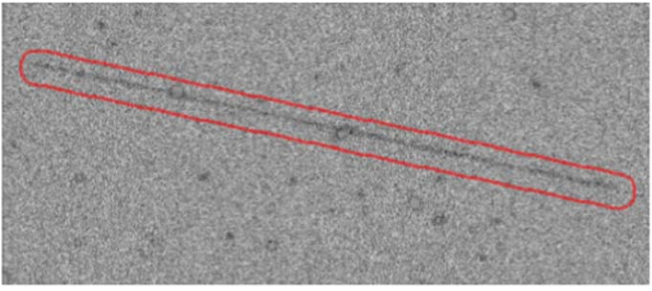

**Figure 1.** A pill aperture drawn around a simulated streak. Notice one of several challenges associated with this technique, the overlap of the target streak with unwanted sources (even those that have been subtracted).

### 2.3.2. Extending PSF Photometry

Another option for photometry is to recover the number of counts by fitting a model for the streak to the streak in the image. We construct our model as follows: assume that the PSF of the observation is known, which we call $\mathcal{P}$. The basic streak model is then the convolution of a mathematical line of length $L$ passing through the image at an angle $\phi$ with the observation PSF. We offset the centroid of the streak with parameters $x_c$ and $y_c$, normalize the model so that it integrates to to the model count number $C$, and complete the model by adding a local sky mean $\mu_{\text{sky}}$. Expressed mathematically, this is

$$M(x, y; x_c, y_c, \phi, L, C, \mu_{\text{sky}}) = C \frac{l(x, y; x_c, y_c) * \mathcal{P}(x, y)}{\int l(x', y'; x_c, y_c) * \mathcal{P}(x', y') dx' dy'} + \mu_{\text{sky}}. \quad (8)$$

In Equation (8), $l$ is the mathematical line corresponding to the streak, the angle parameter $\phi$, and the streak length $L$. The integral is over the entire model. Our approach here is similar to that of W. Fraser et al. (2016) but with an expanded parameter set. The model they implement in their package "Trailed-Image Photometry in Python" (TRIPPy) wisely uses fewer fitting parameters, instead constructing the trailed PSF (TSF), equivalent to the integrand in Equation (8), from known information. This can be appropriate for solar system objects such as asteroids. These objects, typically being much farther away than objects in Earth orbit, move through the field much more slowly. This, coupled with good knowledge of the orbits of these objects, means that the length and orientation of the streak can be predicted with high absolute accuracy, and so the length and orientation of the TSF can be calculated, rather than fit as parameters. Though these parameters can in principle be calculated for objects with well-known Earth orbits as well, the comparatively large motion through the field of view (and therefore longer streaks) we expect for our Earth-orbiting light sources mean that our model count values, especially toward the ends of the streak, are especially sensitive to the streak length and orientation parameters. For this reason, we fit these as parameters here rather than calculating their expected values separately.

### 2.3.3. Fitting Statistics

A typical choice for fitting a fitting statistic to be minimized, $\chi^2$, is not applicable in this situation. $\chi^2$ is only an appropriate statistic when the observed data is normally distributed, since $\chi^2$ is the sum of standard normal statistics $Z_i^2$ for all $i$ bins in a data set. Our pixel counts are Poisson distributed. Even though the Poisson distribution converges to a Gaussian for high enough count numbers, the bias introduced by minimizing using $\chi^2$ can still be significant and will affect the accuracy of our fitting (P. J. Humphrey et al. 2009). The quantity we are interested in computing accurately is the logarithm of the likelihood function $\mathcal{L}$. For our model, the likelihood function (the probability of a specific outcome defined by the set of observed pixel values $O_i$) is

$$\mathcal{L} = \prod_i \frac{E_i^{O_i}}{O_i!} e^{-E_i}, \quad (9)$$

for expected (mean) pixel values $E_i$, which gives a log-likelihood function of

$$\ln(\mathcal{L}) = \sum_i [O_i \ln(E_i) - E_i - \ln(O_i!)]. \quad (10)$$

It is often desirable to simplify this expression and reduce the time it takes for a computer to evaluate it. Specifically, the logarithm of a factorial of a potentially large number is problematic. Stirling's approximation, $\ln(N!) \approx N \ln(N) - N$, can be applied to reach an approximation for $\ln(\mathcal{L})$, which will be very accurate when pixel values are not small:

$$\ln(\mathcal{L}) \approx -\frac{C}{2} \quad || \quad C \equiv 2 \sum_i \left[ (E_i - O_i) + O_i \ln\left(\frac{O_i}{E_i}\right) \right]. \quad (11)$$

$C$ is a statistic known as *Cash's C* (W. Cash 1979). $C$ is defined with the extra factor of 2 so that it converges to $\chi^2$ in the limit of large observed and expected values (compare to $\ln(\mathcal{L})$ for a set of Gaussian bins, $-\chi^2/2$). Since observed pixel values are the sum of source and sky Poisson counts, the sky brightness should be considered as part of the model when computing the likelihood of a set of pixel values. This is very important; since the sky and source counts are Poisson random variables, the sum of which is also a Poisson random variable, it is important that the unsubtracted sky be considered. Subtracting the sky, and then fitting the streak model alone without the sky parameter, will affect the likelihood function and will generally produce different best fits even for the same streak. If the sky were brighter, then the distribution of counts for a single pixel which has contributions from the model and the sky will have greater variance, since the Poisson distribution variance is just its mean value. If we subtract the sky, the distribution variance will remain large, but the mean will have been reduced. The distribution of counts is then neither described by Poisson statistics nor Gaussian statistics, and we cannot minimize $C$ to fit our model. This means that, in contrast to traditional photometry, this approach does not require difficult sky subtraction; it is in fact hindered by it.

### 2.3.4. Fitting Using Markov Chain Monte Carlo

With exact (Equation (10)) and approximate (Equation (11)) expressions for $\ln(\mathcal{L})$, we only need a fitting algorithm. Gradient descent and similar algorithms are not safe to use for optimization in this context, since they are prone to finding local rather than global maxima in the likelihood function. Other algorithms exist for efficiently sampling from distributions with high-dimension parameter spaces. Markov chain Monte Carlo (MCMC) algorithms, such as the Metropolis–Hastings algorithm (W. K. Hastings 1970), are excellent for this purpose. Such algorithms are often used for fitting models with many parameters to data. In this approach, a large number of "walkers" are initialized and are iteratively moved to new states in the parameter space, with the likelihood of a new state being accepted being calculated using the likelihood function. Over enough iterations, the walkers converge to the posterior distribution defined by the likelihood function and thereby allow measurement of the most likely parameters. Here we will be using the Python package *emcee* (D. Foreman-Mackey et al. 2013), an implementation of the efficient affine-invariant ensemble sampler proposed by Goodman and Weare (J. Goodman & J. Weare 2010).

### 2.4. Theoretical Precision

With the procedure for zero-point calculation established, we turn to theoretical uncertainties in our calibration given sufficiently small Gaussian uncertainties in input parameters. For a dependent quantity $f$, which is a function of the set of parameters $x_i$, the $1\sigma$ uncertainty in $f$ is related to the same uncertainty in each $x_i$ by:

$$\sigma_f^2 = \sum_i \left(\frac{\partial f}{\partial x_i}\right)^2 \sigma_{x_i}^2. \quad (12)$$

Using (12), straightforward calculus and Equations (2), (4), and (6) one can derive the following expression for magnitude zero-point variance:

$$\sigma_{M_0}^2 = \left(\frac{2.5}{\ln 10}\right)^2 \left[\left(\frac{\sigma_I}{I}\right)^2 + \left(\frac{\sigma_{F_0}}{F_0}\right)^2 + \left(\frac{\sigma_C}{C}\right)^2\right]. \quad (13)$$

Here $I$, $F_0$, and $C$, respectively, refer to the satellite irradiance at the observatory, the flux zero-point, and the number of counts associated with the streak. From here, one can use (6) to break up the irradiance uncertainty into uncertainties in satellite position, orientation, and engineering specifics, again using (10). We will explore this in several scenarios now. Note that as before, we assume the light source is an integrating sphere, with a cosine intensity profile about its axis.

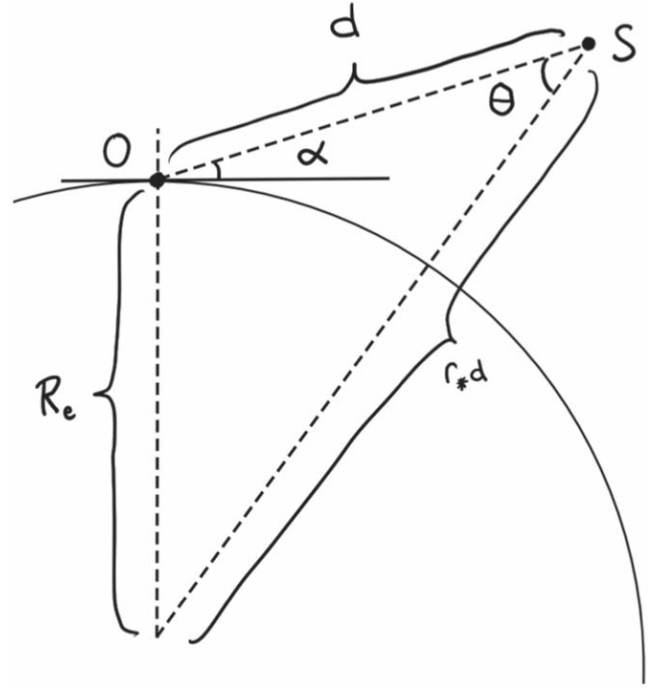

**Figure 2.** Nadir-oriented geometry. The position of the satellite (S) is described in relation to the observation point (O) by its altitude angle $\alpha$ and distance $d$. In this scenario, the axis of the integrating sphere output port is oriented directly toward the center of the Earth.

### 2.4.1. Satellite Oriented to Local Nadir

One natural orientation for the satellite during the observation is to its local nadir (i.e., straight down toward the center of the Earth from the satellite point of view). In this situation, the integrating sphere axis angle $\theta$ can be expressed in terms of the altitude angle $\alpha$ of the satellite; that is, the angle above the horizon which the satellite appears at as seen by the observatory, and the distance to the satellite, $d$. A diagram of this situation is shown in Figure 2. Since the local nadir for the satellite will not be the same absolute direction as the local nadir at the observatory owing to the curvature of the Earth, the radius of the Earth $R_e$ shows up in the resulting expression for $\theta$:

$$\theta = \frac{\pi}{2} - \alpha - \sin^{-1}\left(\frac{\cos(\alpha)}{r_*}\right) \quad \| \quad r_*^2$$
$$\equiv \left(\frac{R_e}{d}\right)^2 + 2\left(\frac{R_e}{d}\right)\sin(\alpha) + 1. \quad (14)$$

Applying Equation (12) to Equations (7) and (14) allows us to express the uncertainty in the irradiance $I$ in terms of $A_{IS}$, $R_{IS}$, $\alpha$ and $d$, and uncertainties thereof:

$$\sigma_I^2 = \left(\frac{\partial I}{\partial R_{IS}}\right)^2 \sigma_{R_{IS}}^2 + \left(\frac{\partial I}{\partial A_{IS}}\right)^2 \sigma_{A_{IS}}^2$$
$$+ \left(\frac{\partial I}{\partial \alpha}\right)^2 \sigma_\alpha^2 + \left(\frac{\partial I}{\partial d}\right)^2 \sigma_d^2 \quad (15)$$

where the partial derivatives are given by

$$\frac{\partial I}{\partial R_{IS}} = \frac{I}{R_{IS}}, \quad \frac{\partial I}{\partial A_{IS}} = \frac{I}{A_{IS}},$$

$$\frac{\partial I}{\partial \alpha} = R_{IS}A_{IS}\sin(\theta)\left[1 - \frac{\frac{\sin(\alpha)}{r_*} + \frac{2}{r_*^2}\left(\frac{R_e}{d}\right)\cos^2(\alpha)}{\sqrt{1 - \left(\frac{\cos(\alpha)}{r_*}\right)^2}}\right],$$

$$\frac{\partial I}{\partial d} = -2R_{IS}A_{IS}\left[\frac{\cos(\theta)}{d^3} + \left(\frac{R_e^2}{d} + R_e\sin(\alpha)\right)\left(\frac{\sin(\theta)\cos(\alpha)}{d^4 r_*^2 \sqrt{1 - \left(\frac{\cos(\alpha)}{r_*}\right)^2}}\right)\right].$$

The above expressions allow one to compute the uncertainty in the magnitude zero point directly from uncertainties in satellite position, orientation, and engineering parameters, in the case that the satellite is oriented to its local nadir.

### 2.4.2. Satellite Oriented to Observation Site

One could also orient the satellite directly toward the observation point at the moment a calibration observation is made. This has two benefits. First, the cosine profile of the irradiance with satellite orientation is maximized when the integrating sphere aperture axis is oriented directly toward the observation point. Second, since the derivative of this cosine function vanishes along this axis ($\theta = 0$), the sensitivity of the irradiance to slightly imperfect orientation also vanishes, and therefore the impact of imprecise alignment is minimized when the satellite is oriented this way. The uncertainty calculations are straightforward compared to the nadir-oriented situation. Equation (11) still applies, but now the variance of the irradiance is given by

$$\sigma_I^2 = \left(\frac{\partial I}{\partial R_{IS}}\right)^2 \sigma_{R_{IS}}^2 + \left(\frac{\partial I}{\partial A_{IS}}\right)^2 \sigma_{A_{IS}}^2 + \left(\frac{\partial I}{\partial d}\right)^2 \sigma_d^2 \quad (16)$$

where the partial derivatives are given by

$$\frac{\partial I}{\partial R_{IS}} = \frac{I}{R_{IS}}, \quad \frac{\partial I}{\partial A_{IS}} = \frac{I}{A_{IS}}, \quad \frac{\partial I}{\partial d} = -2\frac{R_{IS}A_{IS}}{d^3}.$$

It is worth noting that (12) assumes errors are small enough that dependent functions $f$ can be linearly approximated, so quadratic and higher-order error may still be unaccounted for if parameter uncertainties are very large.

## 3. Streaktools

### 3.1. A Description of Streaktools

So far, we have discussed the basic theory of photometric calibration, how a streak created by an artificial light source in orbit can be used to provide this calibration, and several methods for retrieving the number of counts in an image associated with a streak. Now we seek to demonstrate that realistically simulated streaks can be used for calibration by applying this theory. To this end, we have written *Streaktools* for Python, which allows a user to:

1. generate streaks from satellite engineering specifications and position;
2. generate streaks with any user-defined point-spread function;

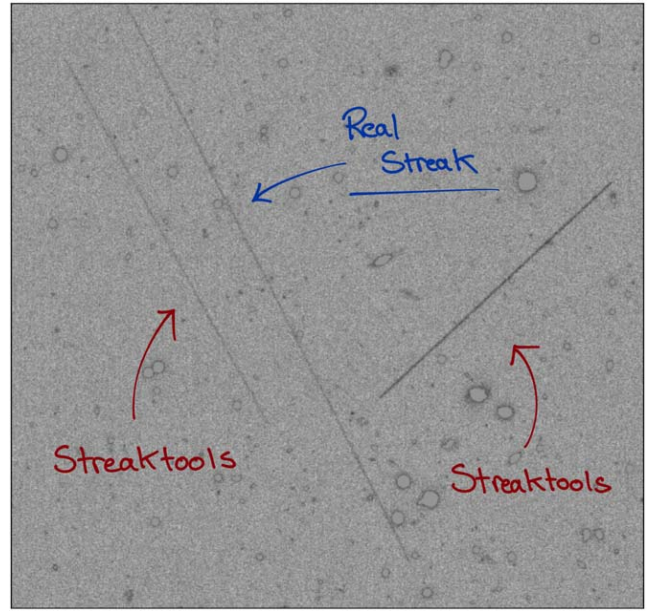

**Figure 3.** A real streak (longest, in the center), and two streaks simulated and added to a DECam image, using Streaktools. These streaks were both generated using the PSF recovered from the image with TRIPPy and SExtractor.

3. add and remove realistic streaks from any real or simulated data;
4. perform pill aperture photometry on simulated streaks;
5. use *emcee* to fit models to data for zero-point recovery;
6. apply the above methods to real streaks.

Shown in Figure 3 is a visual comparison between two simulated streaks generated and added to an exposure of the Dark Energy Camera (DECam) instrument with Streaktools, and a real streak, found in the set of exposures we later use for simulation. Streaktools simulates streaks using a user-defined "true" magnitude zero point, from which the streak is generated. The true magnitude of the streak is calculated using Equation (6), with all of the relevant parameters supplied by the user. After the true number of counts is calculated, the streak model is generated using (8), and an observation of the streak is simulated by computing pixel values as Poisson variables, with mean parameters given by the model at each point. These streaks, with any length, position, and orientation may be added, removed, and changed at will in an image. Results of the photometric methods previously discussed are presented in an easy to read form, automatically calculating and storing the recovered magnitude zero points and best-fit parameters. Methods are implemented in *Streaktools* that tell the algorithm to ignore contributions to the likelihood function calculation beyond a certain distance from the streak, preventing bright nearby sources from skewing likelihood calculations as they are not accounted for in the model. We do not cover the use of Streaktools here; a tutorial is provided with the code which demonstrates all of its features. The code and tutorial may be found publicly accessible on Github[1] and Zenodo[2] (J. L. Goodeve 2024).

## 4. Methods

To test Streaktools and the performance of several calibration methods, we simulated random streaks and added them to data from an exposure taken on 2023 January 31 using the DECam instrument on the Blanco 4 m telescope. These streaks were generated with lengths between 200 and 700 pixels, (clockwise from horizontal) intersection angles from 0° to 90°, randomly positioned in an area covering roughly a quarter of the 2 k × 4 k pixel CCD image used. TRIPPy (W. Fraser et al. 2016) and Source Extractor (E. Bertin & S. Arnouts 1996) were used to recover the PSF of the observation using stars in the image, which was then used for streak simulation and models during calibration. Our aim is to evaluate *Streaktools* and the zero-point recovery methods used, and so satellite source parameters are exactly "known," but are still randomly generated. We generate satellite radiance values between 50 and 1000 W m$^{-2}$ Sr$^{-1}$, a small integrating sphere aperture of 1.27 cm$^2$, generated by a satellite 600km away at an altitude angle of 45° above the horizon. This corresponds to total integrating sphere output of 20 mW to 400 mW. First, 131 streaks were simulated in a sky-subtracted version of the data, and we tested two methods of pill aperture photometry using *Streaktools*:

1. "ordinary" pill photometry: the number of counts inside the aperture is taken to be the final measurement;
2. "compared" pill photometry: the number of counts inside the pill is compared between the exposure with the streak and an exposure from immediately (1 minute) before, and their difference is taken as the measurement of streak flux. This is done in an effort to remove bias from background sources.

---
[1] https://github.com/jgoodeve/streaktools
[2] DOI: 10.5281/zenodo.13941992.

A width parameter $r$ (as detailed in W. Fraser et al. 2016) of 4× the FWHM of the PSF was used for all pills. Next, 425 streaks were simulated in a *non*-sky-subtracted version of the data, and MCMC-based maximum-likelihood fitting was performed using *Streaktools*. For these simulations, we use 40 MCMC walkers, initialized uniformly in region of the parameter space centered on the initializing parameter values for each streak, with a spread intended to represent the accuracy within which an initial guess for the streak model parameters could be provided by a user (several pixels for position, 10 pixels for length, half a degree for angle, and a several $\sigma$ spread for counts and sky background using their Poisson statistics). For all of these simulations, we record generated streak parameters, and calibration results as reported by streaktools with associated uncertainties (estimated from sky noise, total counts and pill area for pill photometry, and from statistics of MCMC results for maximum-likelihood fitting). To generate the streaks we use a preset "true" magnitude zero point of exactly 28.88802, which is very close to the true zero-point for the image established through stellar photometry (chosen for realism). The goal for these photometric methods is to recover this preset zero point accurately.

## 5. Results

### 5.1. Pill versus PSF

We find that both forms of pill aperture photometry described yield poor results, especially for longer streaks and when compared with the MCMC fitting results.

Given in Table 1 are statistics for the pill photometry and MCMC/PSF photometry results (in mmag; "millimag" = $10^{-3}$ mag). Shown for the zero-point error $x$ (the difference between the recovered and true zero point) are the mean $\bar{x}$, the standard deviation $s_x$, and the 1$\sigma$ uncertainty in the mean zero-point error $\sigma_{\bar{x}}$. MCMC/PSF photometry on average recovered the true zero-point 2 orders of magnitude more accurately than the pill methods, and did so consistent with zero bias, unlike the pill results. As intended, the compared pill photometry is not biased toward flux overestimation the way ordinary pill photometry is, but both pill methods' results reflect impractically poor precision. These results are shown in Figure 4, alongside the much more promising results obtained by the MCMC/PSF method.

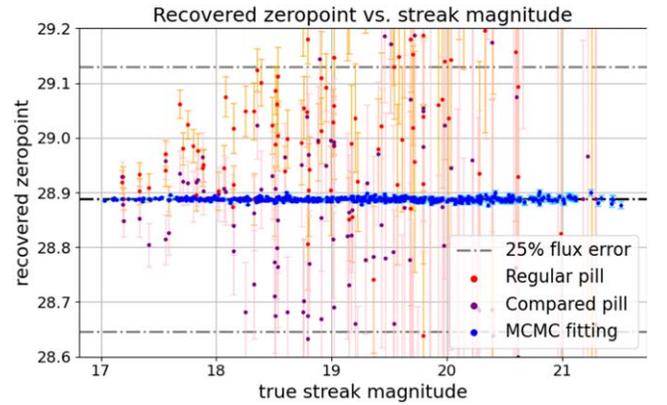

**Figure 4.** Recovered zero point vs. streak magnitude for the three methods.

**Table 1**
Zero-point Recovery Statistics for all three Results

| Parameter | Pill | Compared | MCMC/PSF |
|---|---|---|---|
| $\bar{x}$ (mmag) | −135 | 304 | 0.0308 |
| $s_x$ (mmag) | 486 | 411 | 3.41 |
| $\sigma_{\bar{x}}$ (mmag) | 42 | 36 | 0.165 |

### 5.2. MCMC/PSF Photometry Results

Using MCMC photometry, we were consistently able to recover the image zero point accurately. Over 425 simulated streaks, the mean deviation of the results from the true zero-point was only 0.0308 mmag with a standard deviation of 3.40 mmag. The standard deviation of the mean is therefore 0.165 mmag, and we can therefore say with 99.7% (3$\sigma$) confidence that any systematic bias in these results owing to Streaktools has a magnitude of less than 0.031 mmag under these (fairly typical) circumstances. Shown in Figures 5 and 6 are the zero-point error plotted against two measures of streak brightness; their intensity (in counts per unit length of the streak) and magnitude. The error bars are estimates of the uncertainty in the mean of the posterior distribution. As one might expect, zero-point error is improved with more intense and overall brighter streaks, but this relationship is less strong than for pill photometry. Figure 7 shows a histogram of the full distribution of simulated streaks and their zero-point error using the MCMC approach.

## 6. Discussion

Our results demonstrate (a) that pill photometry applied to light streaks from LEO light sources cannot provide photometric calibration of sufficient accuracy for modern science, but (b) that MCMC/PSF fitting, as described here, could be plausibly used for very accurate photometric calibration that could be a useful and competitive complement or alternative to currently established, ground-based methods. Our 1$\sigma$ zero-point error of 3.41 mmag (flux error ∼0.3%) is comparable to what may be reached by contemporary ground-based photometry (J. D. Hartman et al. 2005; C. W. Stubbs & J. L. Tonry 2006). Our results show several patterns, which can be followed in Figures 5 and 6. First, as for the pill photometry (Figure 4), fainter streaks (quantified by either intensity or magnitude, since source radiance and pulse length both contribute to the total brightness) do generally lead to a broader distribution of results, but this trend is not as strong as for pill photometry. To more conclusively demonstrate the potential of this approach for photometric calibration, more simulations, and especially calibrations using real orbital sources, will need to be performed. This will be necessary to help identify and quantify other sources of error, especially due to atmospheric effects. *Streaktools* can and should be tested in different circumstances. The accuracy provided by streaks that are comparatively very bright or dim could be interesting to investigate, in pursuit of maximum accuracy and the cost effectiveness of any orbital calibrator, respectively. In addition, we identify several nuances specific to our MCMC/PSF approach. First, a model including a sky background term (as ours does) should be fit to a streak in an image that has *not* been sky subtracted, since the sky signal contributes to total noise even when the mean sky value has been removed. Ignoring this would impact the likelihood function (since our pixel count statistics are Poissonian and our noise and mean value are coupled), introducing bias. Also, observations for light streak photometry should consist of minimally short exposures; only enough time to capture a full "pulse" from a source. Extra exposure time beyond this will only add noise. Calibration sources should be engineered for short, bright pulses rather than longer and dimmer ones, since this would allow the exposure time to be brought down and increases streak intensity, as well as minimizing the risk of overlap between the generated streak and contaminating background sources. The only caveat is that exceptionally short pulses may make measuring the image PSF, which is required for our fitting procedure, difficult. This could become an extra source of error.

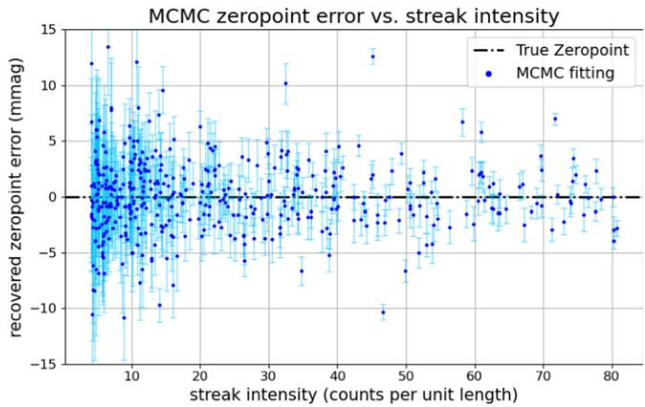

**Figure 5.** MCMC/PSF zero-point error vs. streak intensity, defined as the total number of counts divided by the length, in pixels.

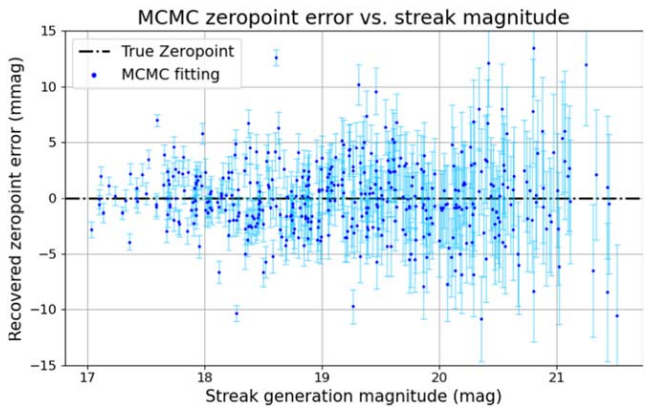

**Figure 6.** MCMC/PSF zero-point error vs. streak magnitude.

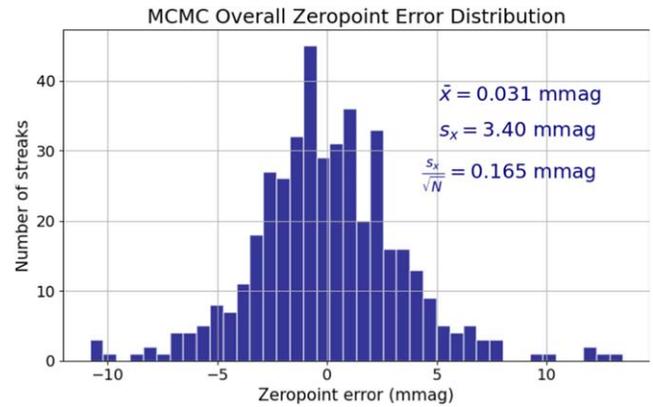

**Figure 7.** MCMC/PSF fitting zero-point error distribution.

## 7. Conclusion

Here we have introduced and explored the concept and theoretical backing of light-streak photometric calibration using artificial light sources in LEO, and have introduced *Streaktools*, publicly available code with which to simulate light streaks and perform light streak photometry. Using *Streaktools*, we have shown pill photometry to be too inaccurate for precise calibration, but that MCMC-based maximum-likelihood model fitting is able to provide an unbiased and accurate ($1\sigma$ error = 3.41 mmag) estimate of the zero point of a real exposure, with streaks simulated using reasonable light-source specifications. This suggests that photometric calibration using LEO sources has strong potential for use in precise photometric calibration, which should be explored.

## Acknowledgments

The research leading to the work presented here was performed at the University of Victoria under the insightful supervision of Dr. Justin Albert, University of Victoria faculty. The author is deeply grateful to Dr. Albert for his encouragement and for the opportunity to work with him on this project.

## ORCID iDs

Joshua L. Goodeve  https://orcid.org/0009-0002-6486-1291